\documentclass[conference]{IEEEtran}

\usepackage[caption=false,font=normalsize,labelfont=sf,textfont=sf]{subfig}

\usepackage{xcolor}
\usepackage{xspace}
\ifCLASSINFOpdf
\else
\fi
\usepackage[cmex10]{amsmath}

\usepackage{balance}
\usepackage{times}
\usepackage{fancyhdr,graphicx,amsmath,amssymb}
\usepackage[linesnumbered,ruled,vlined]{algorithm2e}
\let\oldnl\nl
\newcommand{\nonl}{\renewcommand{\nl}{\let\nl\oldnl}}
\usepackage{algorithmic}

\usepackage{fixltx2e}
\usepackage{dblfloatfix}
\usepackage{makecell}
\usepackage{booktabs}

\usepackage{cite}

\def\bz{{\bf z}}

\def\bh{{\bf h}}

\def\bw{{\bf w}}

\def\ba{{\bf a}}

\def\bs{{\bf s}}

\def\be{{\bf e}}

\usepackage{psfrag}

\begin{document}




\title{Distributed Beamforming in Massive MIMO Communication for a Constellation of Airborne Platform Stations}

\author{\IEEEauthorblockN{\textsuperscript{} Hesam Khoshkbari}
\IEEEauthorblockA{\textit{Department of Electrical Engineering} \\
\textit{École de Technologie Supérieure}\\
Montréal, Canada \\
hesam.khoshkbari.1@ens.etsmtl.ca}
\and
\IEEEauthorblockN{\textsuperscript{} Georges Kaddoum}
\IEEEauthorblockA{\textit{Department of Electrical Engineering} \\
\textit{École de Technologie Supérieure}\\
Montréal, Canada \\
Georges.Kaddoum@etsmtl.ca}
\and
\IEEEauthorblockN{\textsuperscript{} Bassant Selim}
\IEEEauthorblockA{\textit{Department of Electrical Engineering} \\
\textit{École de Technologie Supérieure}\\
Montréal, Canada \\
bassant.selim@etsmtl.ca}
\and
\IEEEauthorblockN{\textsuperscript{} Omid Abbasi}
\IEEEauthorblockA{\textit{Department of Systems and Computer Engineering} \\
\textit{Carleton University}\\
Ottawa, Canada \\
omidabbasi@sce.carleton.ca}
\and
\IEEEauthorblockN{\textsuperscript{} Halim Yanikomeroglu}
\IEEEauthorblockA{\textit{Department of Systems and Computer Engineering} \\
\textit{Carleton University}\\
Ottawa, Canada \\
halim@sce.carleton.ca}
}

\IEEEaftertitletext{\vspace{-2.22\baselineskip}}

\maketitle

\begin{abstract}
Non-terrestrial base stations (NTBSs), including high-altitude platform stations (HAPSs) and hot-air balloons (HABs), are integral to next-generation wireless networks, offering coverage in remote areas and enhancing capacity in dense regions. In this paper, we propose a distributed beamforming framework for a massive MIMO network with a constellation of aerial platform stations (APSs). Our approach leverages an entropy-based multi-agent deep reinforcement learning (DRL) model, where each APS operates as an independent agent using imperfect channel state information (CSI) in both training and testing phases. Unlike conventional methods, our model does not require CSI sharing among APSs, significantly reducing overhead. Simulations results demonstrate that our method outperforms zero forcing (ZF) and maximum ratio transmission (MRT) techniques, particularly in high-interference scenarios, while remaining robust to CSI imperfections. Additionally, our framework exhibits scalability, maintaining stable performance over an increasing number of users and various cluster configurations. Therefore, the proposed method holds promise for dynamic and interference-rich NTBS networks, advancing scalable and robust wireless solutions.
\end{abstract}
\begin{IEEEkeywords}
High-altitude platform station (HAPS), Airborne platform station, Beamforming, Entropy-based multi-agent deep reinforcement learning (DRL)
\end{IEEEkeywords}
\IEEEpeerreviewmaketitle
\section{Introduction}  
Non-terrestrial base stations (NTBSs) are expected to play an instrumental role in next-generation wireless communications, meeting increasing demands and providing widespread connectivity \cite{kurt2021vision,alam2021high}. In this realm, airborne platform stations (APSs), including high-altitude platform stations (HAPSs) and hot-air balloons (HABs), have gained significant attention due to their quasi-stationary positioning, lower latency, and minimal path loss compared to satellites \cite{kurt2021vision,alam2021high,8438489}. Despite these advantages, challenges like user association, beamforming, and interference management arise when using APSs in wireless networks. To address such challenges, the authors in \cite{shamsabadi2022handling} explored power and sub-carrier allocation in HAPS-integrated networks with multiple terrestrial base stations (TBSs) to maximize spectral efficiency (SE). The study in \cite{alsharoa2020improvement} investigated a three-layer network of terrestrial, aerial, and space layers, focusing on the joint link association and power allocation to enhance the network's sum-rate. The authors of \cite{10103832} utilized HAPS as a computing center for mobile vehicles to minimize computation interruptions caused by handovers between roadside units (RSUs). In \cite{10304301}, a HAPS acted as a relay in a terrestrial network (TN) with multiple TBSs, facilitating communication with a geostationary earth orbit (GEO) satellite. This setup optimized user association and beamforming vectors to maximize the network's sum-rate. In addition, \cite{10445467} addressed inter-layer interference between HAPS and TN by developing an iterative algorithm to solve the user association and beamforming optimization to maximize SE. Finally, the authors of \cite{10379023} explored augmenting TN with a cloud-enabled HAPS (C-HAPS) to achieve both global coverage for rural areas and ample capacity for hyper-digitalized zones.

Building on the success of deep reinforcement learning (DRL) in TNs \cite{feriani2021single}, recent research has explored DRL applications in non-terrestrial networks (NTNs) \cite{6928432}. 
In this context, \cite{cao2020deep} introduced a deep Q-learning (DQL) approach for user association in NTNs to maximize the sum-rate while minimizing handoffs due to NTBSs mobility. In our previous work \cite{10171805,10330634,10312746}, we addressed user association in multiple-input multiple-output (MIMO) HAPS-integrated networks to enhance sum-rate. In \cite{10171805}, we proposed a DQL method for user association between HAPS and TBS with only delayed channel state information (CSI) available. This work was expanded in \cite{10330634} to reduce CSI exchange overhead, employing a deep state-action-reward-state-action (SARSA) method where the agent solely relies on delayed terrestrial CSI for user scheduling. In \cite{10312746}, we examined a three-layer network with a TBS, HAPS, and satellite, limiting available CSI to users’ previously associated BS.

Despite numerous studies on resource allocation in APS-integrated networks, distributed beamforming for a constellation of APSs remains underexplored. In this paper, we aim to achieve ubiquitous connectivity by proposing a distributed stochastic beamforming approach using an entropy-based multi-agent DRL for a massive MIMO network with a constellation of APSs, where each BS (i.e., HAPS or HAB) acts as an agent and calculates its beamforming vector using imperfect CSI in both training and testing stages. As highlighted in \cite{10417095}, robust beamforming against imperfect CSI is essential due to jittering in APS placement. Additionally, we account for mobile users and time-varying channels, where rapid channel changes challenge accurate CSI acquisition at each time slot.
Among previous works, only \cite{10171805,10330634,10312746} addressed imperfect CSI HAPS-integrated networks. However, these studies assumed perfect CSI during training, and evaluated imperfect CSI only in the test phase. Furthermore, unlike prior studies that focused on single HAPS and multiple TBS beamforming \cite{10304301,10445467}, this paper investigates beamforming for a constellation of APSs. We assume each APS only has access to its users' imperfect CSI, performing beamforming independently without requiring neighbor APSs' CSIs.
 
\section{system model and problem formulation}\label{sec:system model}
\begin{figure}[h]
	\centering
	\includegraphics[scale=0.14]{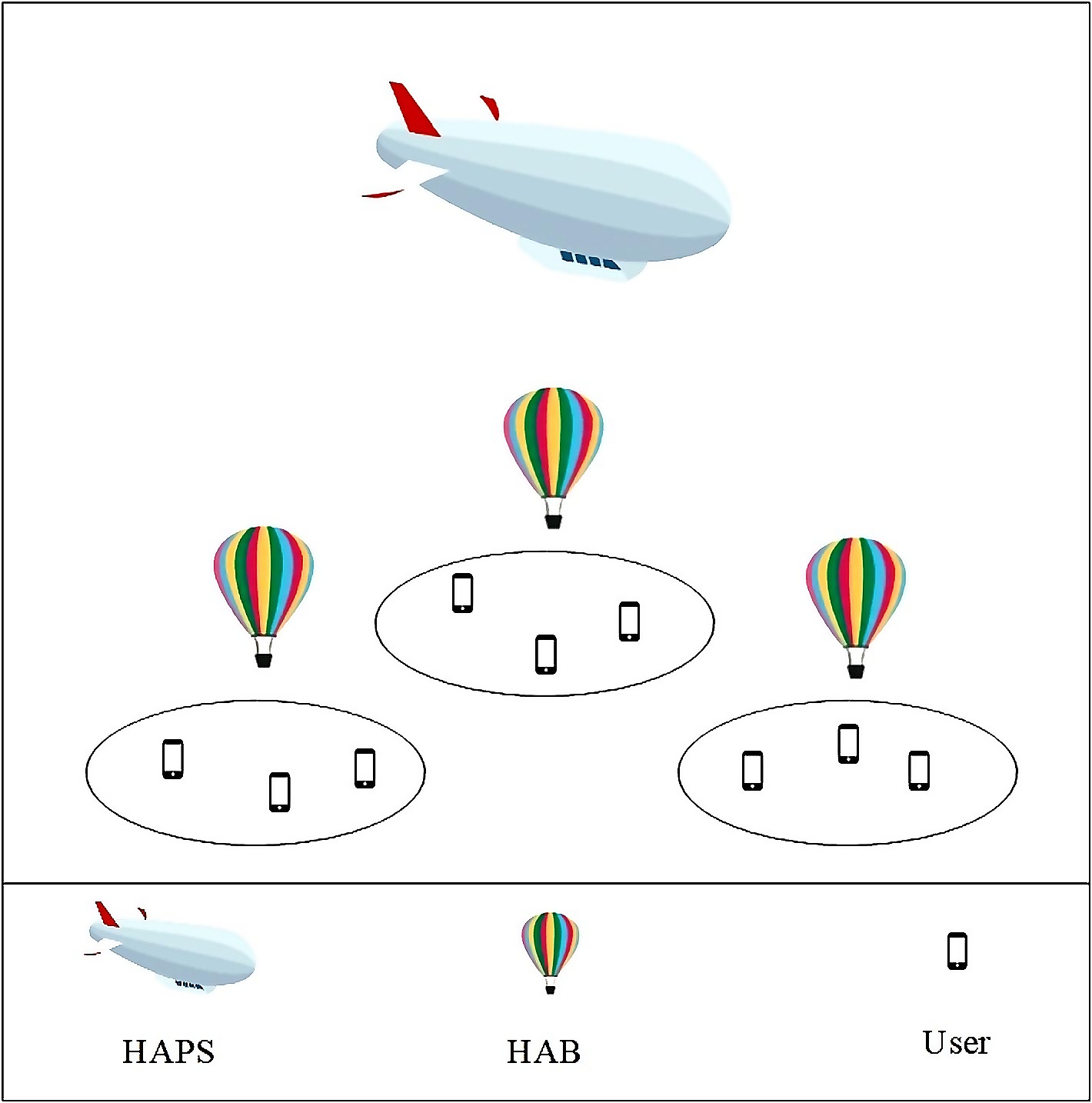}
	\caption{System model illustration. }
	\label{fig:systemmodel}
\end{figure}

This paper proposes a distributed beamforming method for a constellation of APSs to maximize the network’s sum-rate. We consider a two-layered massive MIMO network with a set of APSs $\mathcal{B} = \{b_{0}, b_{1},b_{2},\ldots,b_{B} \}$  (with index 0 reserved for HAPS), each equipped with $N_b$ antennas. The first layer includes $B$ HABs, each serving the users in its own cluster over a shared spectrum. We assume a set of single-antenna users $\mathcal{U} = \{u_{1},u_{2},\ldots,u_{U} \}$, where $U$ = $K \times B$ and $K$ ($N_b \geq K$) represents the number of users per cluster. The second layer consists of a HAPS with $N_{b_0}$ antennas, operating at a separate frequency from the first layer. Dual connectivity—where users are served by two BSs simultaneously—can enhance the sum-rate when line-of-sight (LoS) channels are present \cite{10643015}. Thus, in line with recent studies showing that multi-connectivity improves coverage probability and average achievable data rate in NTNs \cite{10530195}, we assume each user can be served simultaneously by both the HAB in its cluster and the HAPS. Fig. \ref{fig:systemmodel} illustrates our system model for $B  = 3$ HABs and $K = 3$.

At time slot $t$, the channel coefficient between user $u$ and BS $b$ is defined as
\begin{equation}
\bh^{t}_{b,u} = \widehat{\bh}_{b,u}^{t} \sqrt{L_{b,u}^{t}},  
    \label{eq:TBS channel}
\end{equation}
where $\bh_{b,u}^{t}$ is the $1 \times N_{b}$ channel vector between user $u$ and the antennas of BS $b$. Here, $L_{b,u}^{t}$ and $\widehat{\bh}_{b,u}^{t}$ represent large-scale and small-scale fading, respectively. The large-scale fading is defined as
\begin{equation} \label{eq:largescaleTBS}
    \log_{10} L_{b,u}^{t} = \log_{10} \left(\frac{c}{4 \pi f_c d_{b,u}^{t}}\right)^2 - \psi_{d B},
\end{equation} 
where $f_c$ denotes the carrier frequency, $c$ is the speed of light, $d_{b,u}^{t}$ is the distance between user $u$ and BS $b$ at time slot $t$, $\psi_{d B}$  is a Gaussian random variable with zero mean and variance $\sigma_{\psi_{d B}}^2$. Given the LoS nature of APSs, $\widehat{\bh}_{b,u}^{t}$ is derived as
\begin{equation}
  \widehat{\bh}_{b,u}^{t} = \sqrt{\frac{X}{1+X}} \widehat{\bh}_{{b,u}_{LOS}}^{t}+\sqrt{\frac{1}{1+X}} \widehat{\bh}_{{b,u}_{NLOS}}^{t}, \label{eq:LosNLos}
\end{equation}
where $\widehat{\bh}_{{b,u}_{LOS}}^{t}$ and $\widehat{\bh}_{{b,u}_{NLOS}}^{t}$ are the LoS and NLoS components, respectively, and $X$ denotes Rician factor. To account for Doppler effects, we model the small-scale fading $\widehat{\bh}_{b,u}^{t}$ using Jakes' model \cite{dent1993jakes} as
\begin{equation}
\widehat{\bh}_{{b,u}_{NLOS}}^{t} \triangleq   \rho \widehat{\bh}_{{b,u}_{NLOS}}^{t-1} + \sqrt{ 1 -{\rho ^2}} \bz_{u}^{t}, \quad for \quad u = 1 , ..., U, 
\label{eq:HAPS NLOS}
\end{equation}
where $\bz_{b,u}^{t}$ is a complex normal distribution and $\rho$ is the Doppler-dependent correlation factor \cite{abou2005binary}. As noted in \cite{falletti2006integrated} , Jakes' model is suitable for simulating fading in APSs' channels.
The LoS component, $\widehat{\bh}_{{b,u}_{LOS}}^{t}$, is defined as
\begin{equation}
\widehat{\bh}_{{b,u}_{LOS}}^{t} = \mathbf{a}\left(\theta_{b,u}, \phi_{b,u}\right) \otimes \mathbf{b}\left(\theta_{b,u}, \phi_{b,u}\right), \label{eq:HAPS LOS}
\end{equation}
where $\otimes$ denotes the Kronecker product. Moreover, $\mathbf{a}\left(\theta_{b,u}, \phi_{b,u}\right)$ and $\mathbf{b}\left(\theta_{b,u}, \phi_{b,u}\right)$ are defined as
\begin{equation}
\begin{split}
    \mathbf{a}\left(\theta_{b,u}, \phi_{b,u}\right)=\left[1, e^{j 2 \pi d_h}, \ldots, e^{j 2 \pi\left(\sqrt{N_{b}}-1\right) d_h}\right]^T, \\
    \mathbf{b}\left(\theta_{b,u}, \phi_{b,u}\right)=\left[1, e^{j 2 \pi d_v}, \ldots, e^{j 2 \pi\left(\sqrt{N_{b}}-1\right) d_v}\right]^T,
\end{split}
\end{equation}
where $\theta_{b,u}$ and $\phi_{b,u}$ are the elevation and azimuth angles of user $u$ from BS $b$. Moreover, for $\lambda=c / f_{c}$, we have $d_h=d_{\mathrm{x}} \cos \theta_{b,u} \sin \phi_{b,u} / \lambda$ and $d_v=d_{\mathrm{y}} \cos \theta_{b,u} \cos \phi_{b,u} / \lambda$, assuming antenna element spacing $d_{\mathrm{x}} = \frac{\lambda}{2}$ and $d_{\mathrm{y}} = \frac{\lambda}{2}$.

In this paper, as described in Section \ref{sec:method}, we train our proposed framework in an episodic manner, where the environment resets at the start of each episode, lasting $T$ time slots. At the beginning of each episode, each user randomly selects a direction $\Theta_{u} \in [0, 2\pi]$  and moves in that direction with a constant velocity $v$ throughout the episode. Given the coordinates of user $u$ at time slot $t$, ($x_{u}^{t}$,$y_{u}^{t}$), the coordinates at time $t+1$ are
\begin{equation}
\begin{split}
    x_{u}^{t+1} = x_{u}^{t} + D_{\max} \cos \Theta_{u}, \\
    y_{u}^{t+1} = y_{u}^{t} + D_{\max} \sin \Theta_{u},
\end{split}
\end{equation}
where $D_{\max} = v T_{c}$ is the maximum distance a user can travel during a time slot with velocity $v$ and $T_{c}$ denotes the slot duration. We assume, without loss of generality, that each user remains within its assigned cluster during the episode. Thus, if a user reaches the cluster boundary, it selects a new direction and continues moving within the cluster.

We assume that each network layer operates on a distinct frequency band, thus eliminating inter-layer interference. The downlink rate from BS $b$ to user $u$ is defined as
\begin{equation} \label{eq:rate}
    R_{u}^{t} = \log_{2} (1 + \mathrm{SINR}_{b,u}^{t}), 
\end{equation}
where $\mathrm{SINR}_{b,u}^{t}$ is the downlink SINR from BS $b$ to user $u$ at time slot $t$, which for $b \neq b_0$ is given by
\begin{equation}
    \mathrm{SINR}_{b,u}^{t} = \frac{| \bh^{t}_{b,u} \bw_{b,u}^{t}  |^2}{\sum_{\substack{b' \in \mathcal{B} \\b^{\prime} \neq b_0 }} \sum_{\substack{u^{\prime} \\u^{\prime} \neq u} } | \bh^{t}_{b',u} \bw_{b',u'}^{t} |^2+\sigma^{2}}, \label{eq:SINR_TBS}
\end{equation} 
where $\bw_{b,u}^{t}$ is the $N_b \times 1$ beamforming vector from HAB $b$ to user $u$ at time slot $t$. For $b = b_0$, $\mathrm{SINR}_{b,u}^{t}$ is defined as
\begin{equation}
    \mathrm{SINR}_{b,u}^{t} = \frac{| \bh^{t}_{b,u} \bw_{b,u}^{t}  |^2}{\sum_{\substack{u^{\prime} \\u^{\prime} \neq u} } | \bh^{t}_{b,u} \bw_{b,u'}^{t} |^2+\sigma^{2}}. \label{eq:SINR_HAPS}
\end{equation} 

We denote the set of users in cluster $b$ associated with HAB $b$ as $\mathcal{U}_{b}$. For simplicity, we omit the time index and formulate our optimization problem as
\begin{subequations}\label{eq:opt}
\begin{alignat}{4}
\max_{{\bw_{b,u}}} & \quad \sum_{u \in \mathcal{U}} R_{u} \tag{\ref{eq:opt}}\\
\text{subject to} & \quad \sum_{u \in \mathcal{U}_{b}} \left\|\mathbf{w}_{b,u}\right\|_2^2 \leq P_{\max}^{b}, \hspace{0.2cm} \forall b, b \neq b_0 \label{eq:opta}\\
&  \quad \left\|\mathbf{w}_{b,u}\right\|_2^2 \leq P_{\max}^{b}, \hspace{0.2cm} b=b_0 \label{eq:optb} \\ 
&  \quad\quad \cap_{b=1}^{B} \mathcal{U}_{b} = \emptyset \label{eq:optc}\\
& \quad \bw_{b,u} \in \mathbb{C}^{N_b \times 1}, \hspace{0.2cm}  \forall b \label{eq:optd}.
\end{alignat}
\end{subequations}

Here, $P_{\max}^{b}$ represents the maximum power supply at each BS $b$. Constraints \eqref{eq:opta} and \eqref{eq:optb} limit the total allocated power at each BS to its supply power. Constraint \eqref{eq:optc} ensures that each user is only served by the HAB within its own cluster. To solve \eqref{eq:opt},  without access to the perfect and global CSI, we propose a multi-agent entropy-based DRL method, where each BS (HABs and HAPS) functions as an agent to compute the beamforming vector.
\section{Proposed distributed beamforming algorithm}\label{sec:method} 
We assume each agent only has access to imperfect CSI during the training phase. Additionally, mobile users cause $d_{b,u}^{t}$ to vary at each time slot, leading to rapid changes in the channels defined in \eqref{eq:TBS channel}. These assumptions introduce challenges in exploration, as agents require extensive exploration to identify optimal beamforming under imperfect CSI and adapt to the dynamic channels. To address these challenges, we propose an entropy-based multi-agent DRL approach, where each BS acts as an agent to compute its beamforming vector. Each agent includes a stochastic actor that models the policy through a Gaussian distribution, with mean and variance parameters learned by the actor's DNN. The entropy-based exploration encourages more systematic exploration compared to the $\varepsilon$-greedy approach used in \cite{cao2020deep,10171805,10330634,10312746}, allowing the agent to explore different behaviors while avoiding unpromising paths \cite{haarnoja2018soft,haarnoja2018soft2}. In what follows, we first define the state, action, and reward structures and then provide a detailed explanation of our proposed method.
\subsection{Action} \label{sec:action}
The action space for BS $b$ is defined as
\begin{equation}
   \mathcal{A}_{b}= \begin{cases} \{ \Re(\bw_{b,u}), \Im(\bw_{b,u}) \} \hspace{2mm} \text {for} \hspace{2mm} u \in \mathcal{U}_{b}, \forall b, b \neq b_0
    \\   \{ \Re(\bw_{b,u}), \Im(\bw_{b,u}) \} \hspace{2mm} \text {for} \hspace{2mm} u \in \mathcal{U}, b = b_0.\end{cases} 
\end{equation}
where $\Re(\bw_{b,u})$ and $\Im(\bw_{b,u})$ are real and imaginary parts of $\bw_{b,u}$, respectively.

It is important to note that $\bw_{b,u}$  is a complex vector, and since the DNN cannot return complex values, we need to calculate its real and imaginary parts separately.
\subsection{State}
For each BS at time slot $t$, the state is defined as
\begin{equation}
   \bs_{b}^{t}= \begin{cases} [\tilde{\bh}_{b,u}^{t},\bw_{b,u}^{t-1}] \hspace{2mm} \text {for} \hspace{2mm} u \in \mathcal{U}_{b}, \forall b, b \neq b_0,
    \\   [\tilde{\bh}_{b,u}^{t},\bw_{b,u}^{t-1}] \hspace{2mm} \text {for} \hspace{2mm} u \in \mathcal{U}, b = b_0,\end{cases} \label{eq:state}
\end{equation} 
where $\tilde{\bh}_{b,u}^{t}$ is the imperfect CSI, defined as
\begin{equation} 
    \tilde{\bh}_{b,u}^{t}=\xi \bh_{b,u}^{t}+\sqrt{1-\xi^2} \be
\end{equation}
with $\be$ being an error vector with complex normal distribution and $\xi$ representing channel estimation reliability.
Each agent receives the channel coefficients for its associated users only, with no inter-cluster CSI sharing to minimize network overhead. Additionally, each agent receives its previous action, enhancing exploration capability \cite{10330634,10312746}.
\subsection{Reward}
All agents share a common reward, defined as
\begin{equation}
    r^{t} = \frac{1}{U}\sum_{u \in \mathcal{U}} R_{u}^{t}, \label{eq:reward}
\end{equation}
where $R_{u}^{t}$ is the downlink rate of user $u$ defined in \eqref{eq:rate} and $U$ denotes the total number of users. 

\subsection{Proposed entropy-based multi-agent DRL method}\label{sec:method}
\begin{figure}[h]
	\centering
	\includegraphics[scale=0.17]{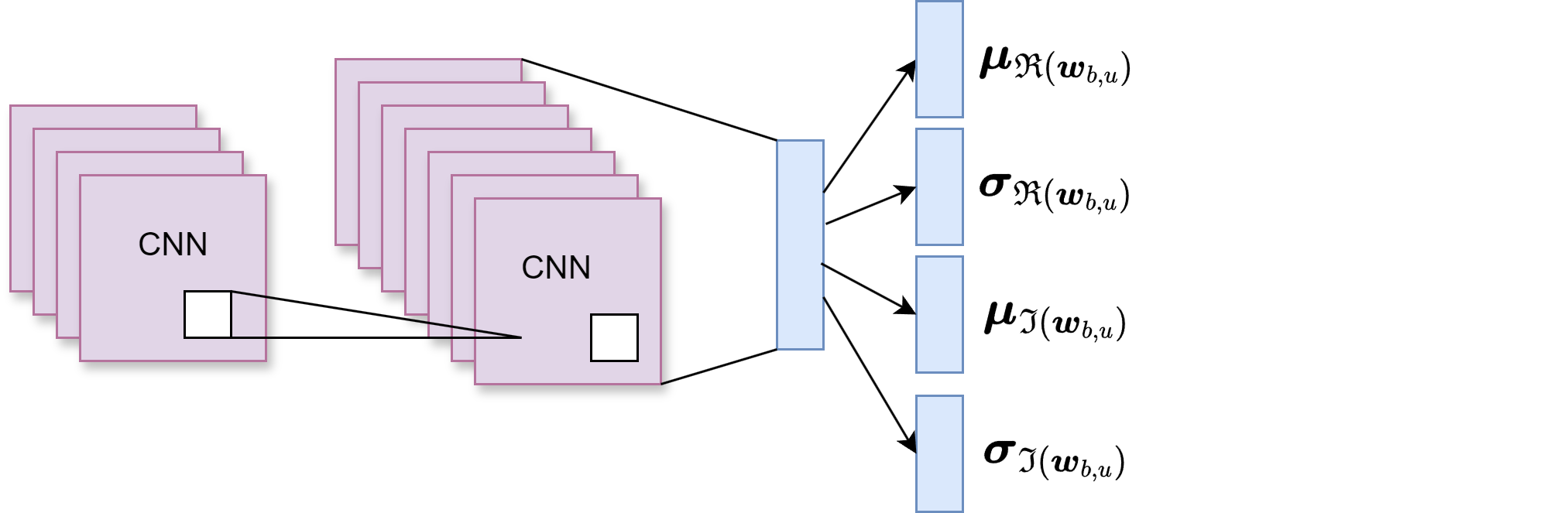}
	\caption{Actor networks diagram. }
	\label{fig:TBS actor}
\end{figure}

In this paper, we use centralized training with distributed execution, training two actor DNNs: the HAB actor network and the HAPS actor network. The beamforming policy for each HAB is modeled by $\pi_{\boldsymbol{\Omega}}\left(\ba_{b}^{t} \mid \bs_{b}^{t}\right)$, while the policy for the HAPS is $\overline{\pi}_{\overline{\boldsymbol{\Omega}}}\left(\ba_{b_0}^{t} \mid \bs_{b_0}^{t}\right)$, with $\boldsymbol{\Omega}$ and $\overline{\boldsymbol{\Omega}}$ being the weights of the HAB and HAPS actors, respectively. During action execution, each BS feeds its state, given in \eqref{eq:state}, into its corresponding actor network to perform beamforming. Fig. \ref{fig:TBS actor} shows the DNN structure; the two networks share the same architecture, differing only in output dimensions and convolutional neural networks (CNN) kernel sizes. The input state in \eqref{eq:state} is reshaped into a matrix with four channels representing the real and imaginary components. Each actor network outputs $\boldsymbol{\mu}_{\Re(\boldsymbol{w}_{b,u})}$, $\boldsymbol{\sigma}_{\Re(\boldsymbol{w}_{b,u})}$, $\boldsymbol{\mu}_{\Im(\boldsymbol{w}_{b,u})}$, and $\boldsymbol{\sigma}_{\Im(\boldsymbol{w}_{b,u})}$. Subsequently, $\Re(\bw_{b,u}) \sim \mathcal{N}(\boldsymbol{\mu}_{\Re(\boldsymbol{w}_{b,u})}, e^{\boldsymbol{\sigma}_{\Re(\boldsymbol{w}_{b,u})}})$ and $\Im(\bw_{b,u}) \sim \mathcal{N}(\boldsymbol{\mu}_{\Im(\boldsymbol{w}_{b,u})}, e^{\boldsymbol{\sigma}_{\Im(\boldsymbol{w}_{b,u})}})$ are sampled , to form the beamforming vector. Thus, $\pi_\Omega\left(\ba_{b}^{t} \mid \bs_{b}^{t}\right)$ and $\overline{\pi}_{\overline{\boldsymbol{\Omega}}}\left(\ba_{b_0}^{t} \mid \bs_{b_0}^{t}\right)$ can be expressed as $f_{\Omega}(\boldsymbol{\epsilon}^{t};\bs_{b}^{t})$ and $\overline{f}_{\overline{\Omega}}(\boldsymbol{\overline{\epsilon}}^{t};\bs_{b_0}^{t})$, respectively, where $\boldsymbol{\epsilon}^{t}$ and $\boldsymbol{\overline{\epsilon}}^{t}$  follow Gaussian distributions with means $\boldsymbol{\mu}_{\Re(\boldsymbol{w}_{b,u})}$ and $\boldsymbol{\mu}_{\Im(\boldsymbol{w}_{b,u})}$ and standard deviations $\boldsymbol{\sigma}_{\Re(\boldsymbol{w}_{b,u})}$ and $\boldsymbol{\sigma}_{\Im(\boldsymbol{w}_{b,u})}$, respectively.

Subsequently, we define the actors' loss functions as
\begin{equation} \label{eq:actor loss}
\begin{split}
J_{\pi}(\boldsymbol{\Omega}) = \mathbb{E}_{(\bs_{b}^{t}, \ba_{b}^{t}, r^{t} ) \sim \mathcal{D}} \left[ \gamma \log(f_{\Omega}(\boldsymbol{\epsilon}^{t};\bs_{b}^{t})) - r^{t}   \right], \\
J_{\overline{\pi}}(\overline{\boldsymbol{\Omega}}) = \mathbb{E}_{(\bs_{b_0}^{t}, \ba_{b_0}^{t}, r^{t} ) \sim \mathcal{D}} \left[ \gamma' \log(\overline{f}_{\overline{\Omega}}(\boldsymbol{\overline{\epsilon}}^{t};\bs_{b_0}^{t})) - r^{t}   \right],
\end{split}
\end{equation} 
where $\mathcal{D}$ is the replay buffer used to store transition data, helping to decorrelate samples and improve training stability \cite{mnih2015human}. Here, $\gamma$ and $\gamma'$  are hyperparameters that control the exploration-exploitation trade-off.

Algorithm \ref{alg:TrainSAC} outlines the training steps for our proposed framework. The input state is reshaped into a matrix with four channels and, as depicted in Fig. \ref{fig:TBS actor}, processed through two CNN layers to extract features. The extracted features are then flattened and passed through a fully connected layer to generate the action via four output layers. We configure the first CNN layer with 4 input channels and 16 output channels. The input and output channels for the second CNN layer is set to 16 and the fully connected layer has 512 units. We initializing weights with a Gaussian distribution. The replay buffer $\mathcal{D}$ is also initialized. All layers use the rectified linear unit (ReLU) activation function, except the output layers. To prevent overfitting to specific network configurations, we reset the environment at the start of each episode by randomly sampling initial channel values from a Gaussian distribution, randomly distributing users within cells, and assigning each user a movement direction. Each BS then receives its respective state matrix and returns beamforming matrices, normalizing these values to satisfy constraints \eqref{eq:opta} and \eqref{eq:optb}. After observing the reward, we store the transition data in the replay buffer $\mathcal{D}$. After each $\eta$ time slot, a mini-batch of data is sampled from the buffer to update the actor networks. Finally, the trained actor networks are saved for evaluation during the test phase.
The testing procedure mirrors training but without weight updates. To evaluate the framework’s performance, we load $\pi_{\boldsymbol{\Omega}}\left(\ba_{b}^{t} \mid \bs_{b}^{t}\right)$ and $\overline{\pi}_{\overline{\boldsymbol{\Omega}}}\left(\ba_{b_0}^{t} \mid \bs_{b_0}^{t}\right)$ obtained from Algorithm \ref{alg:TrainSAC}. Then, as detailed in Algorithm \ref{alg:TrainSAC}, each BS performs beamforming, and we record the network’s sum-rate at each time slot, presenting the average sum-rate in the next section. The episodic evaluation ensures robustness to variations in user locations, movement patterns, and and sampled channel values.
\begin{algorithm}  
  \nonl \textbf{Input}:  replay buffer $\mathcal{D}$, HAB actor network parameters $\boldsymbol{\Omega}$, HAPS actor network parameters $\overline{\boldsymbol{\Omega}}$, system model defined in Section \ref{sec:system model}\;  
  Initialize buffer $D$, and weights $\boldsymbol{\Omega}$, $\overline{\boldsymbol{\Omega}}$.\\
  \For{ $i \in 1 \hspace{2mm} \text{to} \hspace{2mm} I_{\rm{episode}}$}{
      Distribute users, generate the initial channel values and select the movement direction. \\
      \For{$t \in 1 \hspace{2mm} \text{to}  \hspace{2mm} T$}{
           
        Feed state matrix to actor networks and calculate beamforming vectors as explained in Section \ref{sec:method}\;

        Normalize beamforming values to satisfy \eqref{eq:opta} and \eqref{eq:optb}\; 
           
        Observe the reward value \eqref{eq:reward} and store data into $\mathcal{D}$\;
          \uIf{$t$ $\%$ $\eta$= 0}{
              Sample mini-batch of data from $\mathcal{D}$,
             update actor networks using \eqref{eq:actor loss}\;
          }
          }
      \uIf{$i$ $\%$ $\eta'$ = 0}{
          Save the model\;}}
  \nonl \textbf{Output}: $\pi_{\boldsymbol{\Omega}}\left(\ba_{b}^{t} \mid \bs_{b}^{t}\right)$ and $\overline{\pi}_{\overline{\boldsymbol{\Omega}}}\left(\ba_{b_0}^{t} \mid \bs_{b_0}^{t}\right)$\;
\caption{{\color{black}Training algorithm of the proposed entropy-based multi-agent DRL framework}}
\label{alg:TrainSAC}
\end{algorithm}
\section{Simulation Results}\label{sec:results}
During training, we configure $B = 4$ clusters, each with a radius of $q = 2$ km, spaced $l = 6$ km apart. Each cluster contains $K = 4$ users uniformly distributed within its boundaries and one HAB equipped with $N_b = 36$ ($b \neq b_0)$ antennas, operating at an altitude of 2 km. The HAPS, positioned at an altitude of 20 km, is equipped with $N_{b0} = 64$ antennas. During the training, we execute Algorithm \ref{alg:TrainSAC} for $I_{\rm{episode}}$ = 200 episodes, using the Adam optimizer with a learning rate of 0.001 for both the HAB and HAPS actors. The system model, described in Section \ref{sec:system model}, is implemented in Python, with our proposed entropy-based multi-agent framework built using PyTorch. All parameters are listed in Table \ref{tab:sim_param}. Results are averaged over 500 episodes during the test stage.
\begin{table}[h]
\renewcommand{\arraystretch}{1}
\caption{Simulation parameters}
\label{table_example}
\centering
\begin{tabular}{|c||c||c||c|}
\hline
 \textbf{Parameter} & \textbf{Value} & \textbf{Parameter} & \textbf{Value}\\
\hline
 $B$ & 4 & $v$ & 1 m/s \\
\hline
 $N_b$ ($b \neq b_0$) & 36 & $\sigma^{2}$ & -100 dBm\\
\hline
$K$ & 4 & $P_{\max}^{b}$ ($b \neq b_0$)  & 40 watts\\
\hline
$N_{b0}$ & 64 & $P_{\max}^{b_0}$ & 100 watts\\
\hline
$\gamma$ & 0.4 & $\gamma'$ & 0.4\\
\hline
$T$ & 50 & $I_{\rm{episode}}$ & 200\\
\hline
$\sigma_{\psi_{d B}}^2$ & 3 dB & $\overline{\sigma}_{\overline{\psi}{d B}}^2$ & 3 dB\\
\hline
$c$ & 3 $\times 10^8$ m/s & batch size & 32\\
\hline
$\eta'$ & 10 & $\eta$ & 2\\
\hline
$X$ & 10 & $T_{c}$ & 0.02 s \\
\hline
\end{tabular}
\label{tab:sim_param}
\end{table}

\begin{figure}[h]
	\centering
	\includegraphics[scale=0.5]{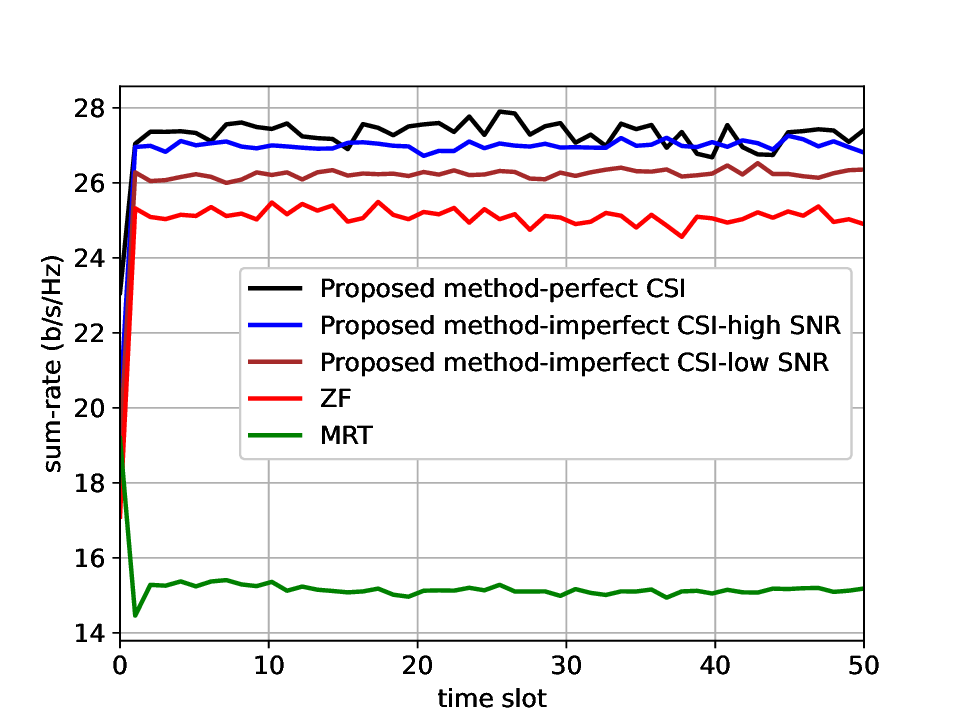}
	\caption{{\color{black}Average sum-rate versus time slot for $K$ = 4.}}
	\label{fig:rate}
\end{figure}

Fig. \ref{fig:rate} shows the average sum-rate at each time slot. We evaluate our proposed method under perfect CSI ($\xi$ = 1), imperfect CSI at high SNR ($\xi$ = 0.8), and imperfect CSI at low SNR ($\xi$ = 0.6), and compare it to the zero forcing (ZF) and maximum ratio transmission (MRT) methods, both evaluated using perfect CSI. It is observed that our method demonstrates strong robustness against imperfect CSI, achieving performances close to that of training with perfect CSI. Compared to ZF, our proposed method achieves an average improvement of 1.9 bps/Hz and 0.14 bps/Hz for $\xi$ = 0.8 and $\xi$ = 0.6, respectively.
This performance improvement can be attributed to the exploration capability of our entropy-based multi-agent DRL method, which performs effectively even under imperfect CSI. Furthermore, as expected, ZF outperforms MRT due to the availability of LoS channels.

\begin{figure}[h]
	\centering
	\includegraphics[scale=0.5]{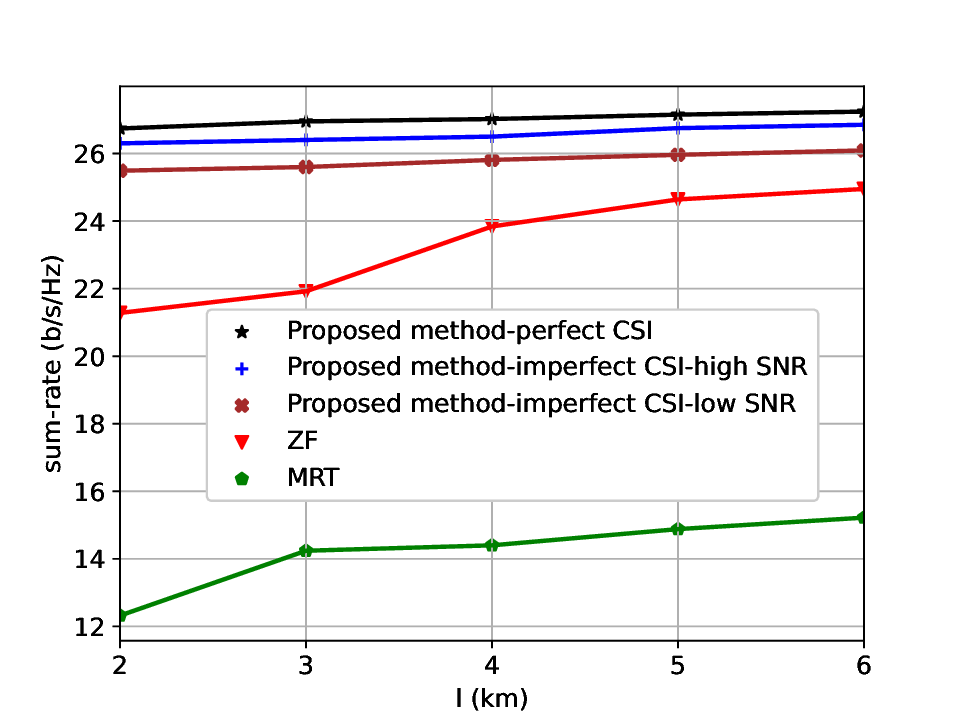}
	\caption{{\color{black}Average sum-rate versus distance between cluster centers.}}
	\label{fig:radius}
\end{figure}

Fig. \ref{fig:radius} illustrates the impact of cluster proximity on user interference and sum-rate performance. As we decrease the distance $l$ (i.e., distance between each two cluster center) to clusters' radius (i.e., $q$ = 2 km), users positioned near the cluster borders experience increased interference from neighboring clusters. Consequently, the sum-rate for ZF and MRT drops as $l$ decreases. Specifically, for the ZF and MRT methods, the sum-rate for $l$ = 2 km drops by 3.67 bps/Hz and 2.9 bps/Hz when $l$ is decreased from $l$ = 6 km, respectively. However, our proposed method effectively selects the optimal users to serve, and mitigate the effects of increasing inter-cluster interference and maintain consistent performance across different $l$ values. It is noted that we used a model trained with $l$ = 6 km and evaluated it across different values for $l$, demonstrating the robustness of our trained model against varying user layouts.

\begin{figure}[h]
	\centering
	\includegraphics[scale=0.5]{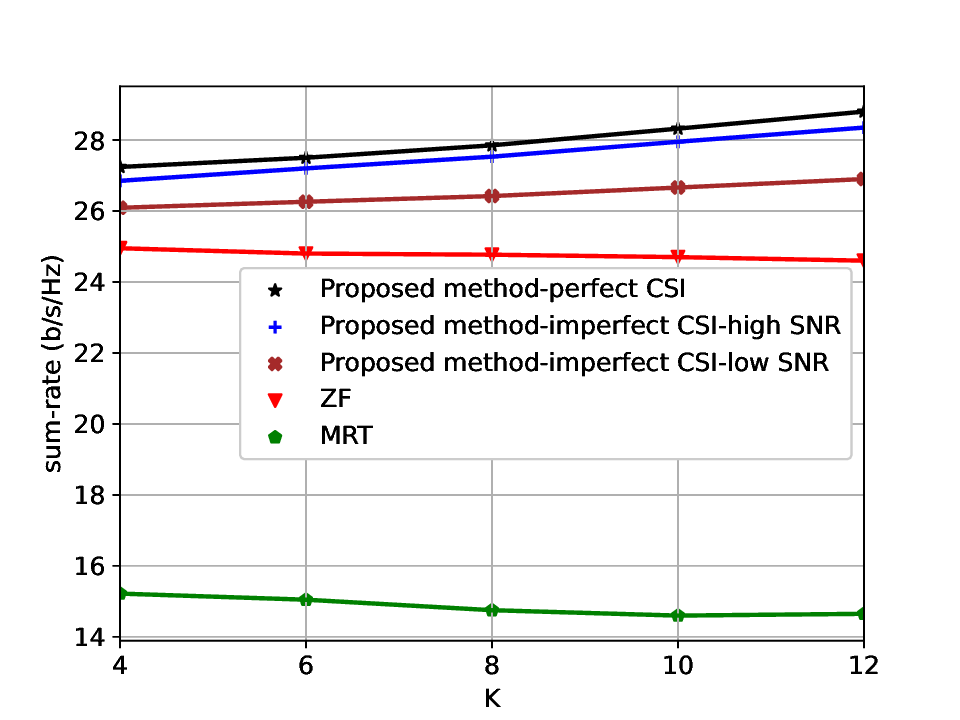}
	\caption{{\color{black}Average sum-rate versus different number of users per cluster.}}
	\label{fig:rate_scale}
\end{figure}

Fig. \ref{fig:rate_scale} shows the average sum-rate for different numbers of users per cluster. Starting from $K$ = 4 to $K$ = 12, the total number of users increases from $U$ = 16 to $U$ = 48. As $K$ increases, user density and intra-cluster interference also rise, expanding the action space and making exploration under imperfect CSI more challenging. However, as shown in Fig. \ref{fig:rate_scale}, our method mitigates the effects of increased intra-cluster interference, resulting in a sum-rate increase of 1.56 bps/Hz for perfect CSI and 1.5 bps/Hz for imperfect CSI at high SNR ($\xi$ = 0.8). Furthermore, the sum-rate gap between our method and the ZF method widens as $K$ grows, due to ZF's deteriorating performance in high-interference scenarios, as previously shown in Fig. \ref{fig:radius}. In contrast, our method remains robust across both low- and high-interference settings, and effectively explore the action space even as $K$ varies, which demonstrates the scalability of our approach.
\section{Conclusions}
This paper presented a distributed beamforming strategy for APS constellations using an entropy-based multi-agent DRL framework in a two-layer massive MIMO network. Our approach effectively addresses challenges posed by imperfect CSI, user mobility, and interference, with each APS operating independently and no CSI sharing among agents. Simulation results validate the robustness of our proposed method, demonstrating its capacity to adapt to various user densities and cluster layouts. Notably, our method consistently outperforms ZF and MRT techniques, maintaining high performance in both low- and high-interference environments. Moreover, the scalability of our framework was confirmed through testing with different user counts and cluster configurations. These findings underscore the potential of our DRL-based method to enhance the sum-rate and resilience of NTBS networks, making it a valuable approach for future large-scale, interference-prone communication systems.




%

\bibliographystyle{IEEEtran}
\balance
\bibliography{myreferences}

@ARTICLE{mnih2015human,
author = {Mnih, Volodymyr and et al.},
journal = {Nature},
title = {Human-level control through deep reinforcement learning},
year = {2015},
volume = {518},
number = {7540},
pages = {529-533},
doi = {10.1038/nature14236}
}

@ARTICLE{10171805,
  author={Khoshkbari, Hesam and Sharifi, Sara and Kaddoum, Georges},
  journal={IEEE Commun. Lett.}, 
  title={User Association in a \uppercase{VH}et\uppercase{N}et With Delayed \uppercase{CSI}: A Deep Reinforcement Learning Approach}, 
  year={2023},
  volume={27},
  number={8},
  pages={2257-2261},
month={Aug.},
  doi={10.1109/LCOMM.2023.3291613}}

@ARTICLE{shamsabadi2022handling,
  author={Alidadi Shamsabadi, Afsoon and Yadav, Animesh and Abbasi, Omid and Yanikomeroglu, Halim},
  journal={IEEE Wireless Commun. Lett.,}, 
  title={Handling Interference in Integrated \uppercase{HAPS}-Terrestrial Networks Through Radio Resource Management}, 
  year={2022},
  volume={11},
  number={12},
  pages={2585-2589},
month={Dec.},
  doi={10.1109/LWC.2022.3210435}}

@article{feriani2021single,
  title={Single and multi-agent deep reinforcement learning for AI-enabled wireless networks: A tutorial},
  author={Feriani, Amal and Hossain, Ekram},
  journal={IEEE Commun. Surveys \& Tuts.},
  volume={23},
  number={2},
  pages={1226--1252},
  year={2021},
  month={2nd Quart.},
  publisher={IEEE}
}

@ARTICLE{falletti2006integrated,
  author={Falletti, E. and Laddomada, M. and Mondin, M. and Sellone, F.},
  journal={IEEE Commun. Mag.}, 
  title={Integrated services from high-altitude platforms: a flexible communication system}, 
  year={2006},
  volume={44},
  number={2},
  pages={85-94},
month={Feb.},
  doi={10.1109/MCOM.2006.1593555}}

@ARTICLE{alsharoa2020improvement,
  author={Alsharoa, Ahmad and Alouini, Mohamed-Slim},
  journal={IEEE Trans. Wireless Commun.}, 
  title={Improvement of the Global Connectivity Using Integrated Satellite-Airborne-Terrestrial Networks With Resource Optimization}, 
  year={2020},
  volume={19},
  number={8},
  pages={5088-5100},
  month={Aug.},
  doi={10.1109/TWC.2020.2988917}}

@ARTICLE{10312746,
  author={Khoshkbari, Hesam and Kaddoum, Georges},
  journal={IEEE Commun. Lett.}, 
  title={Deep Recurrent Reinforcement Learning for Partially Observable User Association in a Vertical Heterogenous Network}, 
  year={2023},
  volume={27},
  number={12},
  month={Dec.},
  pages={3235-3239},
  doi={10.1109/LCOMM.2023.3331216}}

@ARTICLE{10330634,
  author={Sharifi, Sara and Khoshkbari, Hesam and Kaddoum, Georges and Akhrif, Ouassima},
  journal={’ IEEE Open J. Commun. Soc}, 
  title={Deep Reinforcement Learning Approach for \uppercase{HAPS} User Scheduling in Massive MIMO Communications}, 
  year={2024},
  volume={5},
  number={},
  pages={1-14},
  doi={10.1109/OJCOMS.2023.3337044}}

@article{dent1993jakes,
  title={Jakes fading model revisited},
  author={Dent, Paul and Bottomley, Gregory E and Croft, T},
  journal={Electronics Letters},
  volume={13},
  number={29},
  pages={1162--1163},
  year={1993}
}

@article{haarnoja2018soft2,
  title={Soft actor-critic algorithms and applications},
  author={Haarnoja, Tuomas and Zhou, Aurick and Hartikainen, Kristian and Tucker, George and Ha, Sehoon and Tan, Jie and Kumar, Vikash and Zhu, Henry and Gupta, Abhishek and Abbeel, Pieter and others},
  journal={arXiv preprint arXiv:1812.05905},
  year={2018}
}

@inproceedings{haarnoja2018soft,
  title={Soft actor-critic: Off-policy maximum entropy deep reinforcement learning with a stochastic actor},
  author={Haarnoja, Tuomas and Zhou, Aurick and Abbeel, Pieter and Levine, Sergey},
  booktitle={International Conference on Machine Learning},
  pages={1861--1870},
  year={2018},
  organization={PMLR}
}

@ARTICLE{6928432,
  author={Liang, Le and Xu, Wei and Dong, Xiaodai},
  journal={IEEE Wireless Commun. Lett.}, 
  title={Low-Complexity Hybrid Precoding in Massive Multiuser \uppercase{MIMO} Systems}, 
  year={2014},
  volume={3},
  number={6},
  pages={653-656},
  month={Dec.},
  doi={10.1109/LWC.2014.2363831}}

@ARTICLE{cao2020deep,
  author={Cao, Yang and Lien, Shao-Yu and Liang, Ying-Chang},
  journal={IEEE Trans. Commun.}, 
  title={Deep Reinforcement Learning For Multi-User Access Control in Non-Terrestrial Networks}, 
  year={2021},
  volume={69},
  number={3},
  pages={1605-1619},
month={Mar.},
  doi={10.1109/TCOMM.2020.3041347}}

@ARTICLE{kurt2021vision,
  author={Karabulut Kurt, Gunes and et al.},
  journal={IEEE Commun. Surveys Tuts.}, 
  title={A Vision and Framework for the High Altitude Platform Station \uppercase{(HAPS)} Networks of the Future}, 
  year={2021},
  volume={23},
  number={2},
  pages={729-779},
  month={2nd Quart.},
  doi={10.1109/COMST.2021.3066905}}

@ARTICLE{10417095,
  author={Abbasi, Omid and Yadav, Animesh and Yanikomeroglu, Halim and Đào, Ngọc-Dũng and Senarath, Gamini and Zhu, Peiying},
  journal={IEEE Wireless Communications}, 
  title={\uppercase{HAPS} for 6\uppercase{G} Networks: Potential Use Cases, Open Challenges, and Possible Solutions}, 
  year={2024},
  volume={31},
  number={3},
  pages={324-331},
  month={June},
  doi={10.1109/MWC.012.2200365}}

@ARTICLE{10379023,
  author={Alghamdi, Rawan and Dahrouj, Hayssam and Al-Naffouri, Tareq Y. and Alouini, Mohamed-Slim},
  journal={IEEE Trans. on Commun.}, 
  title={Equitable 6\uppercase{G} Access Service Via Cloud-Enabled \uppercase{HAPS} for Optimizing Hybrid Air-Ground Networks}, 
  year={2024},
  volume={72},
  number={5},
  pages={2959-2973},
  month={May},
  doi={10.1109/TCOMM.2023.3348842}}

@ARTICLE{10304301,
  author={Liu, Shasha and Dahrouj, Hayssam and Alouini, Mohamed-Slim},
  journal={IEEE Trans. on Veh. Technol.}, 
  title={Joint User Association and Beamforming in Integrated Satellite-\uppercase{HAPS}-Ground Networks}, 
  year={2024},
  volume={73},
  number={4},
  pages={5162-5178},
  month={Apr.},
  doi={10.1109/TVT.2023.3329168}}

@ARTICLE{10445467,
  author={Shamsabadi, Afsoon Alidadi and Yadav, Animesh and Yanikomeroglu, Halim},
  journal={IEEE Commun. Lett.}, 
  title={Enhancing Next-Generation Urban Connectivity: Is the Integrated \uppercase{HAPS}-Terrestrial Network a Solution?}, 
  year={2024},
  volume={28},
  number={5},
  month={May},
  pages={1112-1116},
  doi={10.1109/LCOMM.2024.3370698}}

@ARTICLE{10530195,
  author={Shang, Bodong and Li, Xiangyu and Li, Zhuhang and Ma, Junchao and Chu, Xiaoli and Fan, Pingzhi},
  journal={IEEE Open J. Commun. Soc.}, 
  title={Multi-Connectivity Between Terrestrial and Non-Terrestrial \uppercase{MIMO} Systems}, 
  year={2024},
  volume={5},
  number={},
  pages={3245-3262},
  month={May},
  doi={10.1109/OJCOMS.2024.3400393}}

@ARTICLE{8438489,
  author={Cao, Xianbin and Yang, Peng and Alzenad, Mohamed and Xi, Xing and Wu, Dapeng and Yanikomeroglu, Halim},
  journal={IEEE J. Sel. Areas in Commun.}, 
  title={Airborne Communication Networks: A Survey}, 
  year={2018},
  volume={36},
  number={9},
  pages={1907-1926},
  month={Sep.},
  doi={10.1109/JSAC.2018.2864423}}

@ARTICLE{10643015,
  author={Saeidi, Mohammad Amin and Tabassum, Hina and Alizadeh, Mehrazin},
  journal={IEEE Trans. on Wireless Commun.}, 
  title={Molecular Absorption-Aware User Assignment, Spectrum, and Power Allocation in Dense \uppercase{TH}z Networks with Multi-Connectivity}, 
  year={2024},
  volume={},
  number={},
  pages={1-1},
  doi={10.1109/TWC.2024.3440888}}

@ARTICLE{10103832,
  author={Ren, Qiqi and Abbasi, Omid and Kurt, Gunes Karabulut and Yanikomeroglu, Halim and Chen, Jian},
  journal={IEEE Trans. on Wireless Commun.}, 
  title={Handoff-Aware Distributed Computing in High Altitude Platform Station (\uppercase{HAPS})–Assisted Vehicular Networks}, 
  year={2023},
  volume={22},
  number={12},
  pages={8814-8827},
  month={Dec.},
  doi={10.1109/TWC.2023.3266344}}

@article{alam2021high,
  title={High altitude platform station based super macro base station constellations},
  author={Alam, Md Sahabul and Kurt, Gunes Karabulut and Yanikomeroglu, Halim and Zhu, Peiying and D`ao, Ngoc},
  journal={IEEE Commun. Mag.},
  volume={59},
  number={1},
  pages={103--109},
  year={2021},
  month={Jan.},
  publisher={IEEE}
}

@article{abou2005binary,
  author={Abou-Faycal, I. and Medard, M. and Madhow, U.},
  journal={IEEE Trans. Commun.}, 
  title={Binary adaptive coded pilot symbol assisted modulation over Rayleigh fading channels without feedback}, 
  year={2005},
  volume={53},
  number={6},
  pages={1036-1046},
  month={June},
  doi={10.1109/TCOMM.2005.849998}}

\end{document}